\documentclass[twocolumn,showpacs,amsmath,amssymb,prl,footinbib,floatfix,superscriptaddress]{revtex4-1}
\usepackage{graphicx}
\usepackage{dcolumn}
\usepackage{bm}
\usepackage{helvet}
\usepackage{amssymb}
\usepackage{amsmath}
\usepackage{amsfonts}
\usepackage{hyperref}
\usepackage{color}

\newcommand{\bra}[1]{\langle#1|}
\newcommand{\ket}[1]{|#1\rangle}
\newcommand{\beq}{\begin{equation}}
\newcommand{\eeq}{\end{equation}}

\newcommand{\sandwich}[3]{\langle#1|#2|#3\rangle}

\newcommand{\eva}[1]{\langle#1\rangle}
\newcommand{\tio}{\mathcal T_{\mathcal C}}
\newcommand{\Tr}{{\rm Tr}}

\newcommand{\contour}{\mathcal C}
\newcommand{\tm}{t_{\rm max}}
\newcommand{\tmm}{t} 

\date{\today}
\begin{document}

\title{First order dynamical phase transitions} 
\author{Elena Canovi}
\affiliation{Max Planck Research Department for Structural Dynamics, University of Hamburg-CFEL, Hamburg, Germany}

\author{Philipp Werner}
\affiliation{Department of Physics, University of Fribourg, 1700 Fribourg, Switzerland}

\author{Martin Eckstein}
\affiliation{Max Planck Research Department for Structural Dynamics, University of Hamburg-CFEL, Hamburg, Germany}

\begin{abstract}
Recently, dynamical phase transitions have been identified based on the non-analytic behavior of the Loschmidt echo in the thermodynamic limit [Heyl et al., Phys.~Rev.~Lett.~{\bf 110}, 135704 (2013)]. By introducing 
conditional probability amplitudes,
we show how
dynamical phase transitions can be further classified, both mathematically, and potentially
in experiment.
This leads to the
definition 
of first-order dynamical phase transitions. 
Furthermore, we develop a generalized Keldysh formalism which allows to use nonequilibrium dynamical mean-field theory
to study the Loschmidt echo and dynamical phase transitions in high-dimensional, non-integrable models.
We find dynamical phase transitions of first order in the Falicov-Kimball model  and in the Hubbard model.
\end{abstract}
\maketitle
The last two decades have witnessed an extraordinary boost in the investigation of strongly correlated systems out of equilibrium,
both experimentally and theoretically. This renewed interest is the consequence of the impressive experimental advances achieved
in the manipulation of cold atoms 
in optical lattices~\cite{Kinoshita2006,Bloch2008a,Joerdens2008a,Schneider2008a}, and in 
ultrafast time-resolved spectroscopy in 
solids~\cite{Iwai03,Perfetti2006,Wall11}.
Using systems of cold atoms, which are very well isolated from the environment and easily tunable,
one can now address fundamental and long-standing problems in statistical physics. 
In particular, many intriguing phenomena have recently been uncovered in relation to the relaxation of excited many-body 
states towards thermal equilibrium \cite{Polkovnikov2011RMP}. 
Thermalization can be hampered due to (near) integrability \cite{Rigol2007}
and delayed by pre-thermalization \cite{Berges2004a,Moeckel2008a,Kollar2011a},
and the different relaxation regimes can be separated by a narrow 
crossover as a function of some parameter
 \cite{Eckstein_PRL09,Schiro11}.
Near 
symmetry-breaking phase transitions, the dynamics can be altered entirely by the presence of non-thermal critical points \cite{Berges2008,Sciolla2010a,Tsuji2012}. 
An unsolved question in this context is whether some of these dynamical crossover phenomena 
reflect an underlying ``sharp'' transition, involving a mathematical non-analyticity of some nature.

In the transverse-field Ising model, Heyl et al.~\cite{Heyl_PRL13} found
a non-analytic time-dependence of the Loschmidt echo,
i.e., the probability to return to the initial state within a non-trivial time-evolution. 
Although the latter is not directly related to the time-dependence of 
thermodynamic 
observables, this
observation 
suggests an intriguing new starting point for analyzing and classifying the dynamical behavior of many-particle systems. To be more precise, we consider a quantum quench, i.e. a sudden change of 
the Hamiltonian from some $H(t<0)=H_0$ to $H(t\geq 0)=H$, which triggers a 
nontrivial out-of-equilibrium evolution. Heyl et al.~\cite{Heyl_PRL13} defined a dynamical phase transition (DPT) 
as 
a non-analytic behavior of the return probability 
amplitude~\footnote{The expression ``dynamical phase transition'' also appears in the literature with another meaning, e.g. in Ref.~\cite{Eckstein_PRL09} it refers to a transition between different relaxation regimes. In this paper we strictly adhere to the definition given  by Heyl et al.~\cite{Heyl_PRL13}.} 
\beq\label{eq:ovampl}
A(t)=\sandwich{\psi_{0}}{e^{-iHt}}{\psi_{0}}
\eeq
as a function of time, where $\ket{\psi_{0}}$ is the ground state of $H_0$.
The return probability, defined by $L(t)\equiv\vert A(t)\vert^{2}$, is the Loschmidt echo. In analogy to the equilibrium partition function, which has a large deviation form $Z = \Tr e^{-\beta H} \sim e^{-\beta N f(\beta)}$ in the thermodynamic limit $N\to\infty$ with a free energy density 
$f(\beta)$, 
$A(t)$ has a large deviation limit of the form $A(t)\sim e^{-Na(it)}$,
and non-analytic behavior as a function of time can occur in the thermodynamic limit
\footnote{Note that differently from the partition function in equilibrium an overlap $A(t)$ can become zero also for finite systems, which would imply a non-analytic behavior of $a(it)$. Such orthogonalities~\cite{Karrasch_PRB13, Andraschko_PRB14,Ligare_AJP02,Giessen_PRA96, Stey_72} usually rely on certain resonances between many-body eigenstates. For a generic finite quantum system $A(t)$ is nonzero for all times, making DPTs a unique phenomenon appearing in the thermodynamic limit.}.

Since the seminal work~\cite{Heyl_PRL13}, further progress has been achieved in the understanding 
of DPTs \cite{Fagotti_pp13,Andraschko_PRB14,Heyl_pp14,Heyl_pp13,Gong_NJP13,Karrasch_PRB13,Vajna_PRB14,Hickey_PRB14,Kriel2014},
but important questions remain open. Firstly,
the Loschmidt echo is the probability of performing no work in a double quench 
experiment $H_0\to H\to H_0$ \cite{Heyl_PRL13},
but 
it is not in any obvious, simple way related to the time-evolution of physical observables, which also hampers a further characterization and classification of DPT's. Furthermore, 
DPTs may be hard to access 
in non-integrable systems which do not allow for an exact solution: 
the computation of an overlap amplitude 
is most direct 
with 
wave-function based numerical techniques, which are however almost exclusively used for
finite or one-dimensional systems.  Examples thereof are  exact diagonalization , 
which is restricted to small systems, or infinite DMRG~\cite{Karrasch_PRB13}.
In this Letter we present  two concepts to address these questions: first we introduce conditional amplitudes and generalized expectation values, which allow for 
a further classification of DPTs and also for the 
definition of first order transitions. Second, we explain how the amplitude \eqref{eq:ovampl} can be computed with diagrammatic many-body techniques and nonequilibrium dynamical mean-field theory \cite{REVIEW}, which makes it accessible for a large class of 
high-dimensional, interacting
models directly in the thermodynamic limit. 

{\em First-order dynamical phase transitions --- }
As Eq.~\eqref{eq:ovampl} gives the 
probability amplitude for the return to the initial state $|\psi_0\rangle$, a natural  way to further classify a DPT is to more closely characterize the ``path'' along which this return happens. As we will see, a first-order 
DPT 
occurs when 
these paths for infinitesimally different propagation times $t$
can be 
distinguished by a nonvanishing change in a {\em macroscopic} measurement.  To illustrate this idea,  let $\hat X\equiv N\hat x$ be any observable  which is extensive in the system size $N$. 
Then we can define a 
{\em conditional return amplitude}
\beq\label{eq:campl}
\tilde A(t,x)\Delta x\equiv \sandwich{\psi_{0}}{e^{-iH(t-t_{1})}\mathcal P^{{\Delta x}}_{x}e^{{-iHt_{1}}}}{\psi_{0}},
\eeq
where $\mathcal{P}^{{\Delta x}}_{x}$ can be any operator that selects eigenstates of $\hat x$ with eigenvalues in a small interval of size $\Delta x$ 
around $x$, e.g.,  $\mathcal{P}^{{\Delta x}}_{x} \propto \sum_{i} e^{-[\langle i | \hat x | i\rangle-x]^2/2\Delta x^2} \ket{i}\bra{i}$. (Note that this choice implies that $\mathcal{P}$, and hence $\tilde A(t,x)$, is a smooth function of $x$ for finite systems).
In a many-body path integral 
formulation \cite{Negele1988a}, Eq.~\eqref{eq:ovampl} can be written as the sum over all paths in some configuration 
space (Grassmann variables for fermions, complex fields for bosons), with a boundary condition provided by the 
state $|\psi_0\rangle$, while $\tilde A(t,x)$ sums the sub-class of paths fixed by the constraint $\hat x=x$ at the intermediate time $t=t_1$.
By construction, we have $A(t)=\int dx\, \tilde A(t,x)$. Assuming again a large deviation form 
$\tilde A(t,x)=e^{-N\tilde a(it,x)}$ 
for $N\to\infty$, the integral will be dominated by its
saddle-point
 values, i.e., $a(it) = \tilde a(it,x_*(t))$, where the complex number $x_*(t)$ is determined by $d\tilde a/dx|_{x=x_*}=0$. 
In the presence of several saddle-points the dominant one can change as a function of the parameter $t$, which defines 
a first order dynamical transition, in analogy to first order transitions in equilibrium. 
Because such a first order transition is a change of the propagator \eqref{eq:ovampl}, its detection should not depend on the particular choice of $\hat x$ or $t_1$, but should be reflected by an abrupt change of the {\em generalized expectation value} of a generic observable $\hat Y$,
\begin{align}
\label{expval_O}
\langle \hat Y(t_1) \rangle _A 
=
A(t)^{-1}\,\sandwich{\psi_{0}}{e^{-iH(t-t_{1})}\hat Ye^{{-iHt_{1}}}}{\psi_{0}} \,.
\end{align}
which is obtained from $A(t)$ by an infinitesimal variation 
$\langle \hat Y(t_1) \rangle _A$=
$i\frac{\delta \ln A_\eta(t)}{\delta \eta(t_1)} |_{\eta=0}$, 
of a field $\eta(t')$ coupling to $\hat Y$, with $A_\eta(t)=\sandwich{\psi_{0}}{T_t \exp[-i\int_0^t dt' (H+\eta(t') \hat Y)]}{\psi_{0}}$.
One of the main results of this work is that both the expectation values Eq.~\eqref{expval_O} and the rate $a(it)$ can be easily computed 
within the  DMFT formalism, as we show later. 
It follows from the discussion above that the  expectation value of $\hat x$ yields the 
complex saddle point $x_*(t)$, 
which 
abruptly changes as a function of $t$.

Before discussing first-order DPTs'  in 
specific models,
it is important to note how generalized expectation values are related to real measurements, in spite of the fact that the quantity $\langle \hat Y \rangle _A $ 
itself is in general complex and only real 
probabilities 
like the Loschmidt echo can be considered measurable. 
To make the connection, we consider the Loschmidt echo, 
$L_{\delta t}(t)\equiv|\sandwich{\psi_0}{e^{-iH(t-t_1)}e^{-ig\hat Y\delta t}e^{-iHt_1}}{\psi_0}|^2$ of an experiment with an extended quench protocol involving a quench $H_0\to H$ at time zero, a short intermediate propagation from $t_1$ to $t_1+\delta t$  with a Hamiltonian 
 $g \hat Y$, and a final propagation with $H$~\footnote{For example, when $Y$ is the double occupancy in the Hubbard model as below, this means a short switch-off of the hopping.}.
Taking the limit of small $\delta t$ yields:
\begin{align}
\label{eq:ratioecho}
L_{\delta t}(t)\,/\,L(t)
=
1+2g\delta t \,\text{Im}\eva{\hat Y}_{A}+\mathcal{O}(g^2\delta t^2).
\end{align}
In essence, the intermediate propagation adds a phase kick to the propagator, thus measuring the imaginary 
part of $\eva{\hat Y}_{A}$. 

{\it Dynamical mean-field theory ---}
We now proceed to explain how the Loschmidt amplitude rate $a(it)$ and the expectation values \eqref{expval_O} can be computed for high-dimensional fermionic lattice systems. 
In the study of quantum systems out of equilibrium, one of the most powerful techniques is dynamical mean-field theory (DMFT)
\cite{Georges96,REVIEW}, which captures local correlations in high-dimensional systems, by mapping a lattice model onto an 
effective impurity model. This mapping is exact in the limit of infinite dimensions \cite{Metzner1989}. Here we 
use it 
to 
study  the generic correlated lattice model 
\beq\label{eq:genham}
H(t)=H_0+U(t)\sum_i n_{i\uparrow}n_{i\downarrow}
\eeq
 with
$
H_0=
-
\sum_{\langle i,j\rangle\sigma}
t_{\sigma}  V_{ij} c^{\dagger}_{i\sigma}c_{j\sigma}
+
\mu
\sum_{i\sigma} n_{i\sigma},
$
which describes fermions with two 
(spin) 
flavors on a lattice:
$V_{ij}$ are lattice-dependent hoppings,
$t_\sigma$ is a spin-dependent prefactor of the hopping term,
and $n_{i\sigma}=c^\dagger_{i\sigma} c_{i\sigma}$. 
The time-dependent  local  repulsion energy $U(t)$ is the parameter driving the sudden quench from $H_0$ to $H$:
$U(t\leq0)=0$  and $U(t>0)=U$. 
The Hamiltonian~\eqref{eq:genham} describes the Falicov-Kimball 
model when one spin flavor is localized ($t_\downarrow=0$), and the Hubbard model when
$t_\uparrow=t_\downarrow=1$ (see below).
 
\begin{figure}[tbp]
\includegraphics[width=0.7\columnwidth]{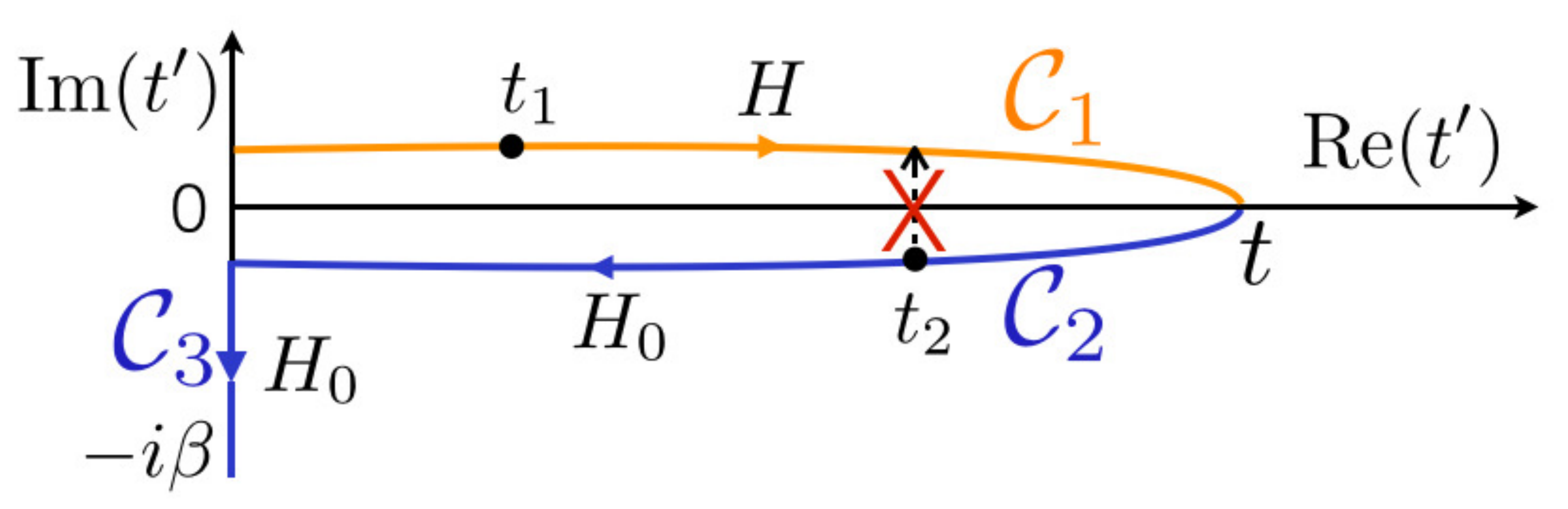}
\caption{(color online) Generalized contour-dependent Hamiltonian on the Keldysh contour $\mathcal C$. The upper, lower and imaginary branches of the contour are denoted by $\mathcal C_{1}$, $\mathcal C_{2}$ and $\mathcal C_{3}$ respectively. The arrows indicate the contour ordering, in this case $t_1$ comes earlier than $t_2$, i.e. $t_1\prec t_2$. For a Green function $G(t_1,t_2)$, if $t_{1}$ lies on $\mathcal C_{1}$ and $t_{2}$ lies on $\mathcal C_{2}$, the latter cannot be shifted to the upper contour. 
}
\label{fig:mokeco}
\end{figure}
 
Nonequilibrium DMFT is based on the many-body Keldysh formalism, which is formulated in terms of Green's functions and 
thus does not directly give access to wave-function overlaps like in Eq.~\eqref{eq:ovampl}. In order to use a Green's function 
formalism to compute the overlap, we first introduce in 
Eq.~\eqref{eq:ovampl} an identity $e^{-iH_{0}\tmm}e^{iH_{0}\tmm}$ and a fictitious temperature $1/\beta$, which is then sent to zero,
\beq\label{eq:ampl1}
A(t)=\lim_{\beta\to\infty}e^{E_{0}(\beta-it)}{\rm Tr}\left(e^{-\beta H_{0}}e^{iH_{0}t}e^{-iHt}\right)\;.
\eeq
Formally, we can view the terms under the trace as the time-ordering of a {\it generalized contour-dependent Hamiltonian} (GCH)
defined on the Keldysh contour $\mathcal C=\mathcal C_1\cup\mathcal C_2\cup\mathcal C_3$,
\beq
\label{ZZZ}
\mathcal{Z}_\mathcal{C}
\equiv
{\rm Tr}\left(\tio e^{-i\int_{\mathcal C}dt'\,H_{\mathcal C}(t')}\right)\;,
\eeq
where the Hamiltonians $H_{\mathcal C}(t)$ on the upper ($\mathcal C_{1}$) and lower ($\mathcal C_{2}$)  real branches are different, $H_{\mathcal C}(t)=H$ for $t\in\mathcal{C}_{1}$ and $H_{\mathcal C}(t)=H_0$  for $t\in\mathcal{C}_{2,3}$ (see Fig.~\ref{fig:mokeco}). We can define contour-ordered expectation values $\eva{\mathcal{O}}_{H_\mathcal{C}} = {\rm Tr} [\tio e^{-i\int_{\mathcal C}dt'\,H_{\mathcal C}(t')} \mathcal{O}(t_1)]/\mathcal{Z}_\mathcal{C}$, which coincide with the generalized expectation values \eqref{expval_O} in the limit $\beta\to\infty$.

At this point we note that the Keldysh formalism
remains applicable
when the Hamiltonian is 
an explicit function of the contour time. 
In particular, diagrammatic rules for contour-ordered Green's functions $G_{ij}(t_1,t_2) = -i\eva{\tio c_i(t_1)c_j^\dagger(t_2)}_{H_\mathcal{C}}$ 
remain unchanged, and one can define a self-energy and a Dyson equation formally identical to those for the standard contour Hamiltonian.
With this, any argument leading to the DMFT formalism, based on either power counting or the cavity formalism \cite{Georges96} can be rewritten one-to-one for a generic contour-dependence of $H_{\mathcal C}$. 
We use DMFT with GCHs to study the Falicov-Kimball and the Hubbard model, in the former using closed equations of motion,
in the latter emplying a Quantum Monte Carlo algorithm \cite{Werner2010}. Details on the DMFT solution
and its implementation are given in the Supplemental Material.
 
Within the Green's function formalism, the overlap amplitude \eqref{eq:ampl1} is obtained  from a coupling constant formalism.
Taking the derivative of the free energy $a_{U}(i\tmm)=\lim_{N\to\infty}-(1/N)\ln A_{U}(\tmm)$ involves the generalized 
expectation value of the double occupancy $d=\frac{1}{N}\sum_in_{i\uparrow}n_{i\downarrow}$,
\beq\label{eq:free-energy-der}
\frac{\partial a_{U}(i\tmm)}{\partial U}=-i\lim_{\beta\to\infty}\int_{0}^{\tmm}dt' \eva{d(t')}_{H_{\mathcal C}(U)}\,,
\eeq
where the dependence of $A$ [Eq.~\eqref{eq:ampl1}] and $H_{\mathcal C}$ on the parameter $U$ in $H$ is made explicit.
For convenience, we  also define the integrated double occupation $\Delta(U,\tmm)\equiv (1/\tmm)\int_0^{\tmm} dt'\, \eva{d(t')}_{H_{\mathcal C}(U)}$.
The free energy is then just the integral of \eqref{eq:free-energy-der}, i.e. $a(it)=\lim_{\beta\to\infty}it\int_0^{U}dU' \Delta(U',t)$.

\begin{figure}[t]
  \includegraphics[width=0.95\columnwidth]{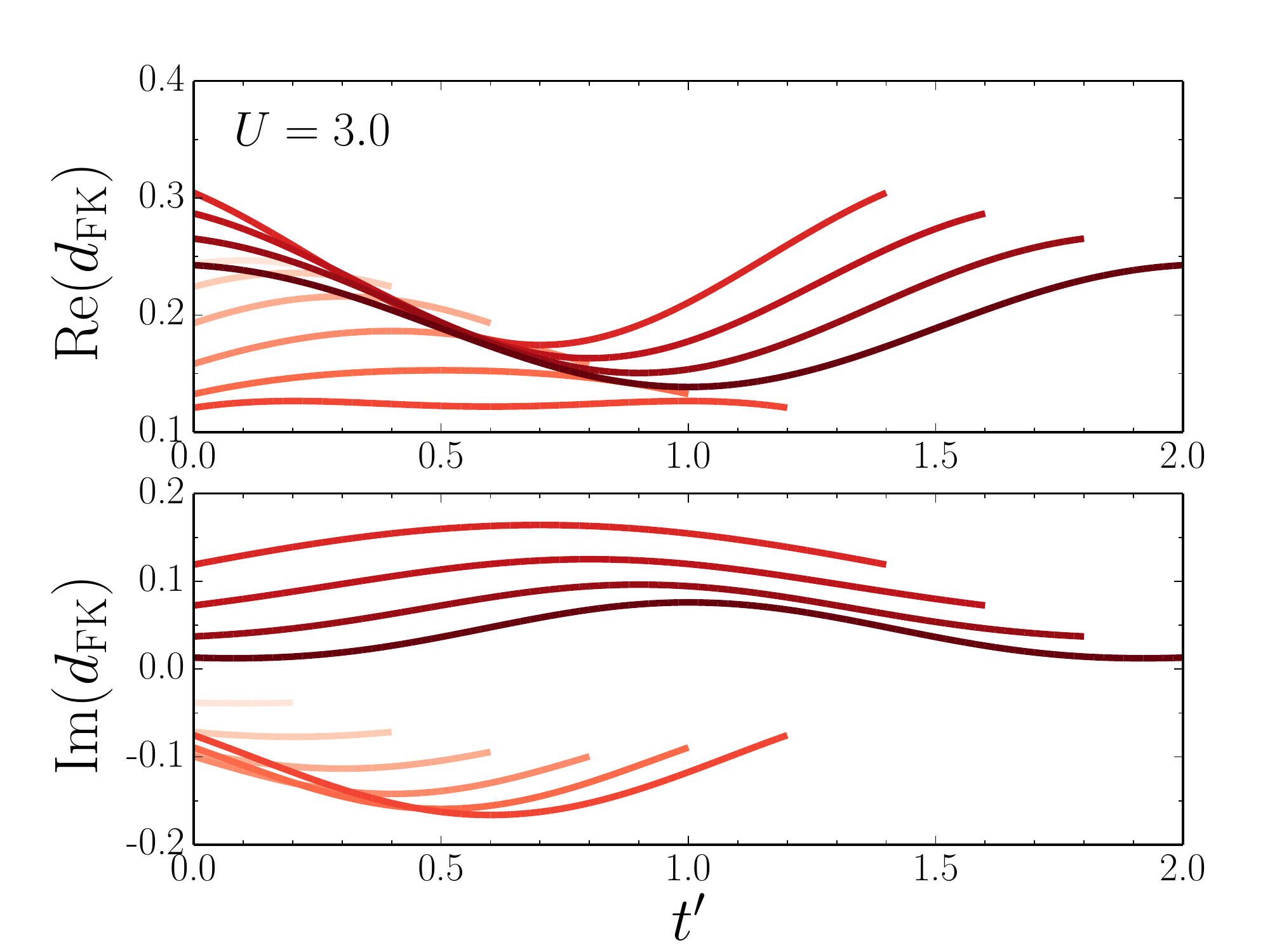}
\caption{ (Color online) 
Time-dependent generalized expectation value of the double occupancy
$d_{FK}(t')\equiv\langle d(t') \rangle_{H_\mathcal{C}}$
in the FKM for $U=3.0$ and increasing values of $\tmm$ 
from $t=0.2$ to $t=2.0$ 
($t$ is evident from the length of the contour, $0\leq t'\leq t$).
Upper panel: real part, lower panel: imaginary part. Data obtained with $\beta=50$, real-time discretization step $dt=0.02$ and a mesh of $N_\tau=200$ points on the imaginary axis (see the Supplemental Material 
for technical details).}
\label{fig:tdocc}
\end{figure}
{\it Results ---}  
As a first application of the above results, we focus on the
Falicov-Kimball model (FKM). 
It describes two species of fermions: the itinerant ones, which can hop between neighboring sites, and the immobile ones,
which act as an annealed disorder potential for the other species. 
The Hamiltonian is given by Eq.~\eqref{eq:genham} with hopping $t_\sigma=0$ for one species.
The FKM can be solved exactly within  DMFT \cite{Brandt1989a}. It displays a rich phase diagram \cite{Freericks2003a}, including a paramagnetic metal-insulator transition at half-filling ($\eva{n_{\uparrow}}=\eva{n_{\downarrow}}=\tfrac12)$ which is located at $U_c=2$ (independent of temperature) for the Bethe lattice.
The possibility of an exact solution makes the FKM an important benchmark also for nonequilibrium DMFT \cite{Freericks2006, Freericks2008, Eckstein_PRL08,Eckstein_NPJ10,Moritz2012}, in spite of the peculiarity that thermalization is excluded because of the missing interaction between the itinerant fermions \cite{Eckstein_PRL08}. 
The DMFT equations for a GCH, which are analogous to the standard nonequilibrium DMFT solution \cite{Freericks2006},
are given in the Supplemental Material.
 
\begin{figure}[t]
  \includegraphics[width=0.95\columnwidth]{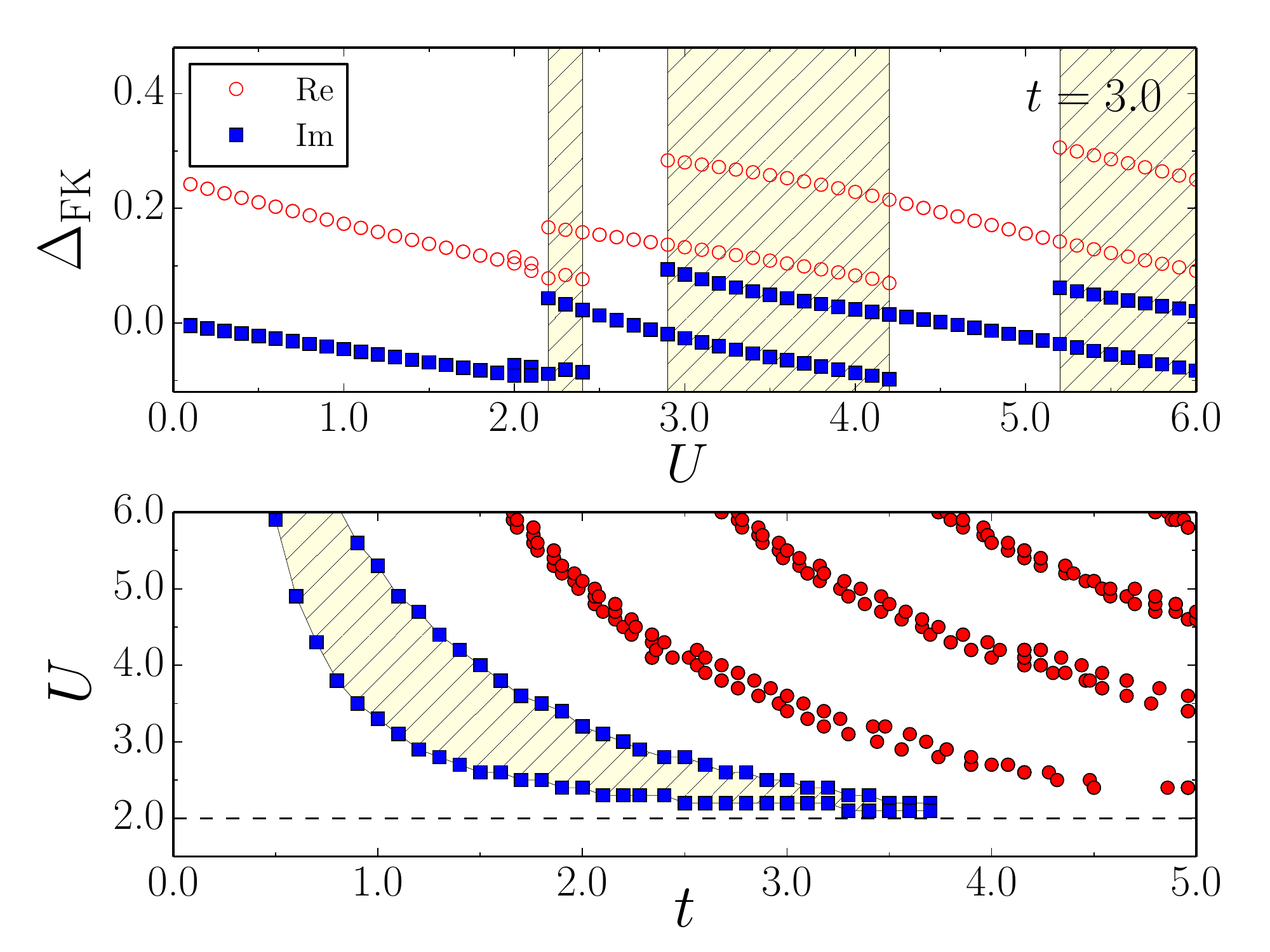}
  \caption{ (Color online) 
Dynamical phase diagram of the FKM.  
Top: real and imaginary part of the integrated double occupation $\Delta_{\rm FKM}$ obtained by increasing (decreasing) $U$ in steps of $\Delta U=0.1$ from $U=0.1$  ($U=6.0$), and using the solution at $U$ as a seed for the iterative solution of DMFT at $U+\Delta U$ ($U-\Delta U$). 
Bottom: 
Blue squares show the coexistence region around the first transition branch, obtained at each $\tmm$ as described in the upper panel. 
For the 
other 
transition branches, we provide
only lower-bound estimates for the coexistence region:
In the 
region between red dots at the same $t$, two coexisting solutions are found by different choices in the 
update of the Green function at each DMFT iteration (see Supplemental Material).
}
  \label{fig:coex}
\end{figure}

We will now show
that the FKM undergoes a DPT. 
As 
can be seen in  Fig.~\ref{fig:tdocc}, our DMFT results indicate that the time-dependent generalized expectation value of the double occupation 
abruptly 
changes its shape with increasing $\tmm$ (see for example the curves at $\tmm=1.2$ and $1.4$). 
In 
Fig.~\ref{fig:coex}
we plot the integrated double occupancy $\Delta(U,\tmm)$ as a function of $U$ for given $\tmm$. 
We indeed find a non-analytic curve, which displays a sequence of jumps 
in whose vicinity
two coexisting DMFT solutions for $d_{\rm FKM}$ are found.
The coexistence of solutions evidences a 
{\em first-order} 
dynamical transition. 
We map out the coexistence region 
(shaded area in the Figure)
by increasing (decreasing) $U$ in small steps, using the solution at a given $U$ as a starting input for the DMFT iteration at the next value of the interaction.  In the lower panel of Fig.~\ref{fig:coex}  blue squares give the bounds of the coexistence region obtained in this way, indicating a transition for quenches to all values $U>U_c$. 
 
We have also applied  
our generalized Keldysh formalism to the Hubbard model, which describes correlated fermions with spin-$\tfrac12$ on a lattice. A numerically exact solution of the nonequilibrium DMFT equations can be obtained with a continuous-time Monte Carlo impurity solver. The weak-coupling approach \cite{Werner2009, Werner2010} allows to simulate reasonably long time intervals, especially in the present set-up, where the time-evolution starts from a non-interacting equilibrium state, and where interaction vertices only have to be sampled on the forward branch $\mathcal{C}_1$. However, 
since the Green functions for GCHs lack causal symmetries (see Supplemental Material),
we cannot use the improved estimator introduced in Ref.~\onlinecite{Werner2010}, which makes the calculations  time-consuming.  

\begin{figure}[t]
  \includegraphics[width=1.0\columnwidth]{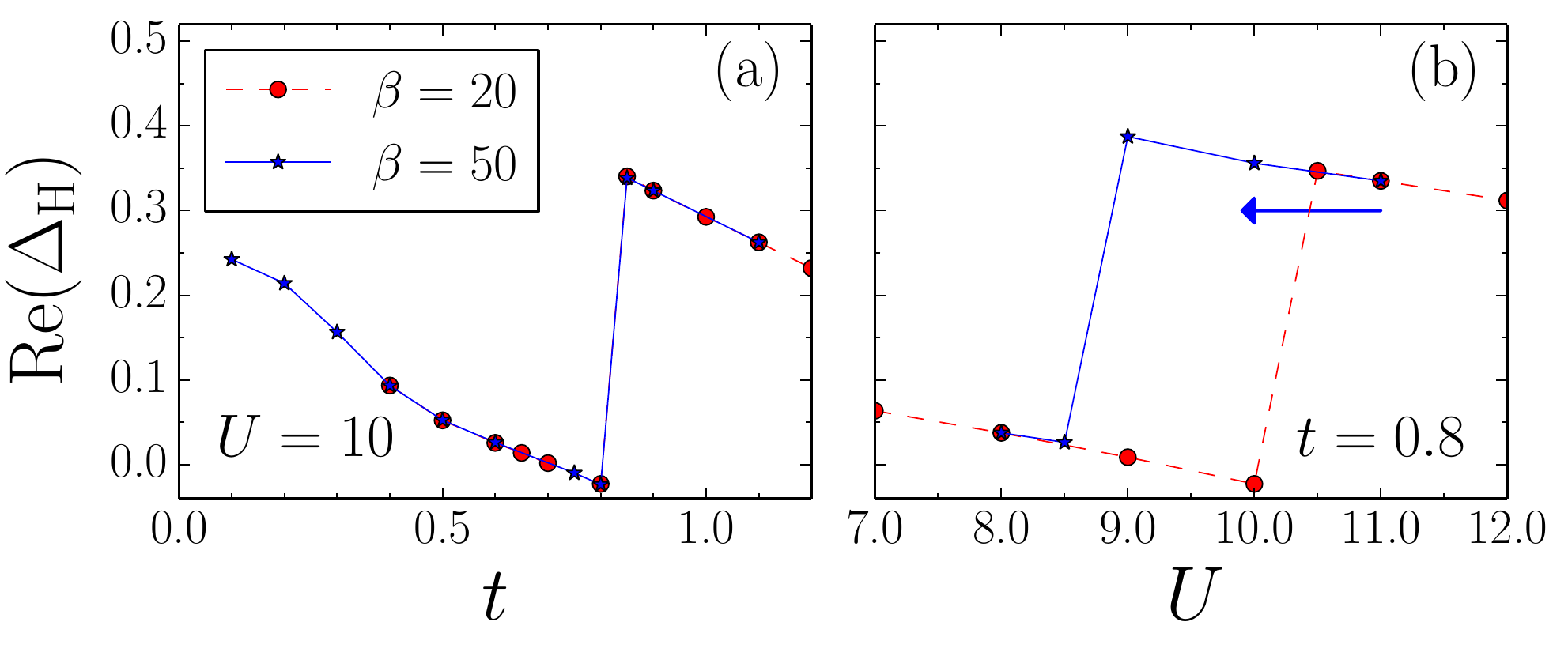}
  \caption{(Color online) DPT in the Hubbard model. Panel (a): Real part of the integrated double occupation for a quench to $U=10$ at different $t$.
To confirm the convergence of the results with the fictitious temperature, we show data for $\beta=20$ (red circles) and $\beta=50$ (blue stars).
Panel (b): coexisting solutions for $\Delta_{\rm H}(U)$ 
at $t=0.8$ at different values of $U$.  Red circles and blue stars are obtained using as an initial guess for the hybridization function in DMFT the noninteracting Green function on the Bethe lattice and  the converged solution at $U=11$  respectively.
}
  \label{fig:hubb}
\end{figure}

Our results 
demonstrate 
that the Hubbard model also exhibits a first-order DPT. 
In 
Fig.~\ref{fig:hubb} (a) we show that the integrated double occupation after a quench in the strong coupling regime ($U=10$)  has a jump at $\tmm\sim0.85$. 
As in the case of the FKM, the first-order nature of the transition is signaled by a coexistence of solutions, as shown in Fig.~\ref{fig:hubb} (b). 
In contrast to the FKM, which is peculiar in the sense that  even in equilibrium the metal-insulator transition prevails to all temperatures, the Hubbard model is a non-integrable model which does show thermalization after a quench \cite{Eckstein_PRL09}.
\par
{\it Conclusions --} This paper provides two
main insights related to the study of DPTs.
From a theoretical point of view, we have shown that dynamical phase transitions can be more deeply characterized 
by means of conditional probability amplitudes
and generalized expectation values, which are experimentally accessible with suitable quench protocols. 
From a methodological point of view, our main result is that the Loschmidt echo can be 
obtained in the context of DMFT  by considering a general contour-dependent Hamiltonian on the Keldysh contour. 
We find first order DPTs both for the Falicov-Kimball and the Hubbard model. 
This raises the hope to actually observe DPT's in experiments with cold atoms, although issues like finite size effects and the influence of the trap remain to be investigated.
In future work we plan to map out the precise phase diagram, including the location of the discontinuities, which requires extensive numerical calculations to perform the additional coupling constant integral.
The presence of first-order DPTs in the FKM and the Hubbard model  can shed new light on the previous works on DPTs. For example, there are indications that the non-analyticity of the Loschmidt rate found in the Ising model ~\cite{Heyl_PRL13}  and its nonintegrable variants ~\cite{Karrasch_PRB13, Kriel2014} are of first-order: the analytical expressions in  Ref.~\cite{Heyl_PRL13}  show that the generalized expectation value of the transverse magnetization $M$, which is a derivative of the Loschmidt rate with respect to the magnetic field, shows a jump at the critical times. (More recent work on 2-band systems indicates transitions of different order \cite{Vajna_pp14}). An intriguing problem would thus be to compute also conditional amplitudes~\eqref{eq:campl} as a function of time and $M$ in this exactly solvable model, and thus to characterize the analytical structure of the transition in this model.

\acknowledgments
We thank K. Balzer, R. Fazio, M. Heyl, S. Kehrein, M. Kollar, J. Mentink, D. Rossini, and S. Sayyad  for useful discussions.
The QMC calculations used a code based on ALPS \cite{Albuquerque2007}.  
PW is supported by FP7/ERC starting grant No. 278023.

\newpage
\appendix
\begin{center}
{\bf Supplemental Material}
\end{center}
\section{Green's functions for a contour-dependent Hamiltonian}\label{app:num}
In the main text we have introduced the 
Keldysh formalism for a generalized contour-dependent Hamiltonian (GCH)
which is different on the two real-time branches of the Keldysh contour $\mathcal{C}$
($H_0$ on the upper and $H$ on the lower branch, respectively.)
As already mentioned, the diagrammatic rules remain unchanged if the Hamiltonian depends explicitly on the contour branch,
and one can define Green's functions $G$, self-energies $\Sigma$, and a Dyson equation formally identical to the 
standard nonequilibrium case,
\begin{equation}
G=G_0+G_0*\Sigma*G
=G_0+G*\Sigma*G_0.\label{eq:dyson}
\end{equation}
($G_0$ is the noninteracting Green's function, and the $*$-symbol denotes the convolution along $\mathcal{C}$.)
In spite of the formal analogy, there are important differences concerning the symmetry of the Green's functions, which have to be taken into account in numerical manipulations. In this section we explain these differences and give details of the numerical implementation of contour convolutions and the inversion of the Dyson equation.

\subsection{Contour-ordered Green's functions}\label{ssec:green}

Contour-ordered expectation values for a GCH are defined in analogy to the standard nonequilibrium formalism
(see Ref.~\cite{REVIEW} for an introduction to the Keldysh formalism and for the notation used in in this text),
\begin{align}
\langle \cdots \rangle_{ H_\mathcal{C}} \equiv \frac{\text{tr} [\tio e^{-i\int_\mathcal{C} dt' H(t')} \cdots] }{\text{tr}[ \tio e^{-i\int_\mathcal{C} dt' H(t')}]}.
\end{align}
Here the contour-ordering is defined as usual, by
\beq
\tio \mathcal A(t)\mathcal B(t')\equiv\theta_{\mathcal C}(t,t')\mathcal A(t)\mathcal B(t')
\pm
\theta_{\mathcal C}(t',t)\mathcal B(t')\mathcal A(t)\;,
\eeq
where 
the upper (lower sign) is for bosonic (fermionic) operators, and
\beq
\theta_{\contour}(t,t')=\left\lbrace
\begin{array}{ll}
1 &t\succ t'\\
0&\text{ else}\;,
\end{array}
\right.
\eeq
with $t\succ t'$ ($t\prec t'$) meaning that $t$ comes later (earlier) than $t'$ in the sense of the contour.
Green's functions are defined as
\beq\label{eq:green}
G(t,t') = -i\eva{\tio c(t)c^\dagger(t')}_{H_\mathcal{C}}\,.
\eeq
As in the standard nonequilibrium case, cyclic invariance of the trace implies 
a boundary condition of the Green's functions,
\begin{eqnarray}
G(0^+,t)&=\pm G(-i\beta,t)\\
G(t,0^+)&=\pm G(t,-i\beta)\;,
\end{eqnarray}
where $0^+\in \mathcal C_1$ and $-i\beta\in\mathcal C_3$.

Because each of its times arguments $t$ and $t'$ can lie on three different branches, the Green's function \eqref{eq:green} has 
9 components $G(t,t')\equiv G_{ij}(t,t')$ ($t\in\mathcal C_i$, $t'\in\mathcal C_j$, $i,j=1,2,3$):
\beq\label{eq:Gmatrix}
\hat G=\left(
\begin{array}{ccc}
G_{11}&G_{12}&G_{13}\\
G_{21}&G_{22}&G_{23}\\
G_{31}&G_{32}&G_{33}
\end{array}
\right)\;.
\eeq
In the standard nonequilibrium case, these $9$ components are not independent from each other: one can always shift the operator with the largest real-time argument from $\mathcal C_1$ to $\mathcal C_2$ and vice-versa, because the backward and forward time evolution operator for larger times cancel. 
Various (time-propagating) approaches found in the literature~\cite{Koehler1999,Tran2008,Stan2009,Eckstein_PRB10,Balzer2013} exploit these symmetries, to transform contour-equations into causal time-propagation equations (Kadanoff-Baym equations). However, these symmetries are apparently lost when the Hamiltonian on $\mathcal{C}_1$ and $\mathcal{C}_2$ is different, so that $G_{11}(t,t')\neq G_{12}(t,t')$ 
even for $t\leq t'$,
 $G_{13}(t,\tau')\neq G_{23}(t,\tau')$, and so on. To manipulate Green's function for a GCH we thus do not use Kadanoff-Baym equations, but stick to an approach based on the explicit discretization of $\mathcal{C}$ (similar to what has been used in Ref.~\cite{Freericks2008} for the standard nonequilibrium case).
\subsection{Discretization of the contour}\label{ssec:discretization}
Each of the real-time branches of $\mathcal{C}$ is divided in $N_t$ intervals, equally spaced with a time step $\Delta t$.
The results presented in this work are obtained with $\Delta t=0.02$. 
Introducing the convention that time on the upper (lower) real contour is indicated with $t^+$ ($t^-$),
the discretized points are $\lbrace t^+_0=0, t^+_1=\Delta t,\dots,t^+_{N_t -1}=(N_t-1)\Delta t,t^+_{N_t}=N_t \Delta t=\tm\rbrace$ and 
$\lbrace t^-_{N_t}=N_t \Delta t=\tm, t^-_{N_t-1}=(N_t-1)\Delta t,\dots,t^-_{1}=\Delta t,t^-_{0}=0\rbrace$ 
on $\mathcal C_1$ and $\mathcal C_2$ respectively.
The point $t=\tm$ is thus  present twice on the discretized contour, so that there are totally $2N_t+2$ points on the real branches.  

Since we have to take the limit of large $\beta$ to study dynamical transitions, particular care has to be taken in the discretization of the imaginary branch. We will take advantage of the fact that the Green's functions on the  imaginary branch vary rapidly only
near $\tau\gtrsim0$ and $\tau\lesssim\beta$, while they vary slowly elsewhere, 
 and 
employ  a nonlinear mesh such that 
the regions $\tau\gtrsim0$ and $\tau\lesssim\beta$ are more densely sampled than the rest of the interval.
 To this aim,  we first  define a mapping from a variable $x\in[0,1]$
to the imaginary time  $\tau\in[0,\beta]$ via a function $f(x)$:
\beq
\tau\equiv\beta f(x)\;,
\eeq
 with a 
 positive 
 derivative $f'(x)$.
The linear mesh is trivially recovered with the function $f(x)=x$.
In our calculations we choose a hyperbolic-tangent mesh defined as follows:
\beq
f(x)=c_1+\frac{c_2}{2}\left(\tanh[\alpha (2x-1)]+1\right)\;,
\eeq
where 
$c_1=-\frac{\rho}{1-2\rho}$, 
$c_2=\frac{1}{1-2\rho}$ and $\rho=\frac{1}{2}(1-\tanh(\alpha))$.
With this definition $f(0)=0$ and $f(1)=1$. We typically take $\alpha=4.0$ to ensure a sufficiently steep function.
We then discretize the variable $x$, splitting the interval $[0,1]$ into $N_\tau$  equally spaced points $x_0=0,x_1=\Delta x,\dots,
x_{N_\tau-1}=(N_\tau-1)\Delta x,x_{N_\tau}=N_\tau \Delta x$, with $\Delta x=1/N_\tau$. 
The nonlinear mesh is now composed of the $N_\tau+1$ points
$\lbrace\tau_n=f(x_n)=f(n \, \Delta x)\rbrace$.
On the discretized contour the $t^-=\tau=0$ point is doubly defined: as $t^-_0=0$ on $\mathcal C_2$ and as $\tau_0=0$ on $\mathcal C_3$.

Summarizing, the 
contour 
is composed of $N_1=N_t+1$ equally space points on $\mathcal C_1$, $N_2=N_t+1$ equally spaced points on $\mathcal C_2$
and $N_3=N_\tau+1$ 
inhomogeneously spaced 
points on $\mathcal C_3$, giving a total of $N\equiv 2 N_t+N_\tau+3$ points.
%
\subsection{Matrix form of the Green's function}\label{ssec:greenfunction}
With the discretization of time described above, the Green's function is represented as a
$N\times N$ matrix $\bar G(t_n,t_m)$, with $n,m=0,\ldots, N-1$.
Since Green's functions are discontinuous at equal times, we store an additional $N$-component vector $\Delta G(t_n)$, which takes into account the 
discontinuity. In our implementation, the diagonal element $\bar G(t_n,t_n)$ contains the average of the lesser and greater components, 
while the vector contains the difference:
\beq\label{eq:Gdiagmat}
\bar G(t_n,t_n)\equiv\frac{1}{2}(G^{<,\rm real}(t_n,t_n)+G^{>,\rm real}(t_n,t_n))
\eeq
and 
\beq\label{eq:Gdiagvec}
\Delta G(t_n)\equiv\frac{1}{2}(G^{<,\rm real}(t_n,t_n)-G^{>,\rm real}(t_n,t_n))\;.
\eeq
The superscripts 
mean the following: 
$<,{\rm real}$ ($>,{\rm real}$) refers to the Green's function $G(t,t')$ with $t<t'$ ($t>t'$) in the sense of {\it real} time (analogously with $\tau < \tau'$ ($\tau > \tau'$) if
the imaginary time is defined with $-i\tau$ with $0\leq\tau\leq\beta$). This does {\it not} coincide
with lesser (greater) in the sense of the contour if, for example, $t$ and $t'$ lie on the lower $\mathcal C^-$ real branch.

We also  introduce a convenient rescaling of the Green's functions on the imaginary branch which incorporates the nonlinear mesh. It consists in 
multiplying the Green's function by a factor $\sqrt{\beta f'}$ for each imaginary time:
\begin{align}
G(t,\tau)&\to \tilde G(t,x)\equiv\sqrt{\beta f'(x)}G(t,\tau(x))\label{eq:rescaling}\\
G(\tau,t)&\to \tilde G(x,t)\equiv\sqrt{\beta f'(x)}G(\tau(x),t),\label{eq:rescaling2}\\
G(\tau,\tau')&\to \tilde G(x,x')\equiv\sqrt{\beta f'(x)}\sqrt{\beta f'(x')}G(\tau(x),\tau(x')).\label{eq:rescaling3}
\end{align}
As we shall see below, the advantage of this rescaling is that the convolution has the same numerical implementation
on the entire contour. 
%
\subsection{Convolutions}\label{ssec:convolution}
The fundamental operation to be implemented for Green's functions is the convolution, which is formally defined as an
integral on the contour: 
\beq\label{eq:convolution}
C(t,t')\equiv[A*B](t,t')\equiv\int_{\mathcal C}d\bar t \; A(t,\bar t)B(\bar t,t'). 
\eeq
From the numerical point of view, two issues must be taken into account in the computation of the convolution:  the nonlinear mesh on the imaginary branch and the effect of the jump of the Green's functions $A$ and $B$ at $t=t'$ . 

Concerning the nonlinear mesh, let us consider the convolution $C(t,t')$ of two Green's functions $A(t,t')$ and $B(t,t')$ (Eq.~(\ref{eq:convolution})) and split it into the three contributions
coming from the different branches $\mathcal C_1, \mathcal C_2$ and $\mathcal C_3$:
\beq\label{eq:3comp}
C(t,t')=C_1(t,t')+C_2(t,t')+C_3(t,t')\;.
\eeq
The rescaling of Eqs.~(\ref{eq:rescaling})-(\ref{eq:rescaling3}) allows to write the convolution in terms of the (equally spaced) variable $x$ in a
straightforword way. As an example, we compute the contribution on the imaginary branch to the  Matsubara component of $C$:
\beq\label{eq:conv_app}
\begin{split}
C_{3}(\tau,\tau')&=\int_0^\beta\, d\bar\tau A(\tau,\bar\tau)B(\bar\tau,\tau')
\end{split}
\eeq
Performing a change of integration variable $\bar\tau=\beta f(\bar x)$ and using 
Eq.~(\ref{eq:rescaling3}), 
we can
rewrite Eq.~(\ref{eq:conv_app}) as
\beq
C_{3}(\tau,\tau')=\int_0^1 d\bar x\, \beta f'(\bar x)\frac{\tilde A(x,\bar x)\tilde B(\bar x,x')}{\sqrt{\beta f'(x)}\sqrt{\beta f'(x')}\beta f'(\bar x)}\;,
\eeq
which implies 
\beq\label{eq:conv_simple}
\tilde C_3(x,x')=\int_0^1 d\bar x\, \tilde A(x,\bar x) \tilde B(\bar x,x')\;.
\eeq
From Eq.~(\ref{eq:conv_simple}) we see that integration on the imaginary branch can be performed on the equally-spaced
variable $x$ at the cost of a simple rescaling (Eqs.~(\ref{eq:rescaling})-(\ref{eq:rescaling3})) of the Green's functions.

In order to clarify how to include the contributions of the jumps of $A(t,t')$ and $B(t,t')$,
we need first to specify how the integrals are discretized on the contour. We found it convenient and efficient to use the trapezoidal rule, 
so that we can generically write $\bar C$ as the result of a matrix multiplication:
\beq\label{eq:intmatrix}
\bar C= \bar A \bar W \bar B\;,
\eeq
where $\bar A$ and $\bar B$ are defined as in Eq.~\eqref{eq:Gdiagmat} and $\bar W$ is a diagonal matrix
$(\bar W)_{nm}=\delta_{nm}w_n$
containing the integration weights. The vector $\bar w$ is composed of three parts, each corresponding to a different branch:
\beq\label{eq:weights1}
\bar w=\left(\begin{array}{c}
\bar w^{(1)}\\
\bar w^{(2)}\\
\bar w^{(3)}
\end{array}
\right),
\eeq
where 
\begin{eqnarray}
w^{(1)}_0=w^{(1)}_{N_t}=\frac{1}{2}\Delta t\,,&\;w^{(1)}_{j}=\Delta t \text{ if } j\neq 0,N_t, \label{eq:weights2}\\
w^{(2)}_0=w^{(2)}_{N_t}=-\frac{1}{2}\Delta t\,,&\; w^{(2)}_{j}=-\Delta t \text{ if } j\neq 0,N_t, \label{eq:weights2b}\\
w^{(3)}_0=w^{(3)}_{N_\tau}=-i\frac{1}{2}\Delta x\,,&\; w^{(3)}_{j}=-i\Delta x \text{ if } j\neq 0,N_\tau,\hspace{3mm}\label{eq:weights2c}
\end{eqnarray}
and in the third line we have used the form (\ref{eq:conv_simple}) of the integral on the imaginary branch.
At this point it is important to remark that $C(t,t')$ itself has no jumps, i.e. $\Delta C(t_n)=0$, so the only thing we have to compute
is the matrix $\bar C$. However, Eq.~(\ref{eq:intmatrix}) alone is not correct, because it does not take into account the discontinuities
of $A$ and $B$ on the diagonal. For this reason, we need to compute corrections to Eq.~(\ref{eq:intmatrix}).
 Suppose we want to compute the diagonal element $C(t,t)$ (with $t\in\mathcal C_1$), and  in particular the contribution of the upper real branch, i.e. $C_1(t,t)=\int_0^{\tm}d\bar t\, A(t,\bar t)B(\bar t,t)$  (see Eq.~(\ref{eq:3comp})).
Using Eq.~(\ref{eq:convolution}), the integral can be exactly rewritten as:
\beq\label{eq:split1}
\begin{split}
C_1(t,t)=&\int_0^{t}d\bar t\, A^{>,\rm real}(t,\bar t)B^{<,\rm real}(\bar t,t) \\
&+ \int_t^{\tm}d\bar t\, A^{<,\rm real}(t,\bar t)B^{>,\rm real}(\bar t,t). 
\end{split}
\eeq
With the discretization described above 
and the integration weights Eqs.~(\ref{eq:weights1}) and ~(\ref{eq:weights2}) , the 
discretized form of Eq.~(\ref{eq:split1}) reads 
\beq\label{eq:intexact}
\begin{split}
(C_{1})_{nn}&=\left\lbrace\sum_{l=1}^{n-1}\bar A_{nl}\bar B_{ln}+\sum_{l=n+1}^{N_t-1}\bar A_{nl}\bar B_{ln}\right.\\
&+\frac{1}{2}\bar A_{n0}\bar B_{0n}+\frac{1}{2}\bar A_{n N_t}\bar B_{N_t n}\\
&+\left.\frac{1}{2} A_{nn}^{>,\rm real} B_{nn}^{<.\rm real}+\frac{1}{2} A_{n n}^{<,\rm real} B_{n n}^{>,\rm real}\right\rbrace\Delta t,
\end{split}
\eeq
where $t=n\Delta t$ and  we used the short notation $\bar A(t_n,t_m)\equiv \bar A_{nm}$. 
The only approximation in Eq.~(\ref{eq:intexact}) with respect to Eq.~(\ref{eq:split1}) is the discretization of time, which in this
case gives an error $\propto\Delta t^2$.
What we actually  find from Eq.~(\ref{eq:intmatrix}), 
using  the representation Eq.~(\ref{eq:Gdiagmat})
of the diagonal elements, is different:
\beq\label{eq:intwrong}
\begin{split}
(\tilde C_{1})_{nn}&=\left\lbrace\sum_{l=1}^{n-1}\bar A_{nl}\bar B_{ln}+\sum_{l=n+1}^{N_t-1}\bar A_{nl}\bar B_{ln}\right.\\
&+\frac{1}{2}\bar A_{n0}\bar B_{0n}+\frac{1}{2}\bar A_{n N_t}\bar B_{N_t n}\\
&+\frac{1}{4} A_{nn}^{<,\rm real} B_{nn}^{<,\rm real}+\frac{1}{4} A_{n n}^{>,\rm real} B_{n n}^{<,\rm real}\\
&+\left.\frac{1}{4} A_{nn}^{<,\rm real} B_{nn}^{>,\rm real}+\frac{1}{4} A_{n n}^{>,\rm real} B_{n n}^{>,\rm real}\right\rbrace\Delta t\;,
\end{split}
\eeq
where we have used explicitly Eq.~(\ref{eq:Gdiagmat}).
Therefore, it is necessary to add a correction 
\beq\label{eq:deltacorr}
(\delta(A*B))_{nn}=(C_{1})_{nn}- (\tilde C_{1})_{nn}\equiv(\Delta C_1)_{nn} 
\eeq
to  Eq.~(\ref{eq:intwrong}) in order to recover Eq.~(\ref{eq:intexact}):
\beq\label{eq:corrections}
\begin{split}
(\delta(A*B))_{nn}=&\left(\frac{1}{4} A_{nn}^{>,\rm real} B_{nn}^{<,\rm real}+\frac{1}{4} A_{n n}^{<,\rm real} B_{n n}^{>,\rm real}\right.\\
&-\left..\frac{1}{4} A_{nn}^{<,\rm real} B_{nn}^{<,\rm real}-\frac{1}{4} A_{n n}^{>,\rm real} B_{n n}^{>,\rm real}\right)\Delta t\\
=&-\Delta A_{nn}\Delta B_{nn}\Delta t\;.
\end{split}
\eeq
Importantly, if we didn't  include the corrections ~(\ref{eq:corrections}), using Eq.~(\ref{eq:intwrong}) instead of Eq.~(\ref{eq:intexact}) in our numerical implementation, 
this would imply an error $\propto\Delta t$. 

Corrections similar to Eq.~(\ref{eq:corrections}) are necessary not only in the diagonal elements  of $C(t,t)$, but also 
every time at least one of $t,t'$ is $0,\tm,\beta$, i.e. boundary terms in $t$ and/or $t'$. For example, in computing $C(0,t)$ the exact integral
contains $A^{<,\rm real}_{00}\bar B_{0n}$, while we compute $\bar A_{00}\bar B_{0n}$.  
Combining all the boundary terms, 
a total of 72 different corrections are needed, plus the already discussed corrections for the diagonal terms. 
\subsection{Inversion and solution of the Dyson equation}\label{ssec:inversion}
The Dyson equations (\ref{eq:dyson}) can be easily recast  in the integral form 
\beq\label{eq:regularized}
(1+F)*Y=C\;,
\eeq
where $F=-G_0*\Sigma$, $C=G_0$ and $Y=G$
is the unknown function.
The DMFT equations for both the FKM and the Hubbard model, described later in the text, can be expressed in the form of Eq.~(\ref{eq:regularized}).

Equations in the regularized form of Eq.~(\ref{eq:regularized}) can be  numerically solved in a straightforward manner using our matrix representation of the Green's functions (see Eqs.~(\ref{eq:Gdiagmat}) and (\ref{eq:Gdiagvec})).
To this aim,  we need first to compute the jump of $Y$, and second, 
find the corrections to $Y$ due to the correction $\delta (F*Y)$ (see Eq.~(\ref{eq:deltacorr})) coming from the convolution.
For the first point, we  note that the diagonal elements of $\bar Y(t_n,t_n)$ and $\Delta Y(t_n)$ are 
{\it independent} variables which in principle satisfy different equations. 
We now observe that $C$ has a jump, while $(1+F)$ has not because it contains the result of a convolution.  
Equation~(\ref{eq:regularized}) then implies
\beq\label{eq:diag}
\Delta Y(t_n)=\Delta C(t_n)
\eeq
for the vector of the differences.\\
If we knew $Y$, we would be able to compute
the convolution $F*Y$  with  all the corrections. On the other side, if none of the functions $F$, $Y$ 
or $C$ had a jump, Eq.~\ref{eq:regularized} would be a simple linear system:
\beq\label{eq:linform}
\left(1+\bar F\bar w\right)\bar Y_0=\bar C,
\eeq
where $\bar w$ is the vector of the integration weights.
We now show that we can find $\bar Y$ solving equations of the form~(\ref{eq:linform}) and using Newton iteration.
Indeed we can view the left-hand side of Eq.~(\ref{eq:regularized}) as a functional $\mathcal F$ of $Y$ and write
\beq\label{eq:dyne}
\mathcal F(Y)=C\;,
\eeq
with
\beq
\mathcal F(Y)=(1+\bar F\bar w)\bar Y+\delta (F*Y)\;,
\eeq
where $\bar F\; \bar w \; \bar Y$ is the contribution to the convolution coming only from the matrix $\bar Y$
and $\delta(F*Y)$ contains the corrections.
As a starting guess for $\bar Y$ we take $\bar Y_0$, i.e. the solution of Eq.~(\ref{eq:linform}).
Notice that we already
know the exact solution for $\Delta Y$ (see Eq.~(\ref{eq:diag})), but for convenience we also define  $\Delta Y_0\equiv\Delta Y$.
 Knowing $\bar Y_0$ and $\Delta Y_0$
we can now compute $\mathcal F[Y_0]$, which contains the full convolution $F*Y_0$, i.e. also the contribution of $\Delta Y_0$. 
Next, we need the derivative of $F$ in $\bar Y_0$, which can be
approximated as
\beq
\left.\frac{d\mathcal F}{d\bar Y}\right|_{\bar Y=\bar Y_0}\approx (1+\bar F\bar w).
\eeq
If we define $\bar Y_1$ as the first correction to $Y_0$, the difference $\delta \bar Y_1\equiv \bar Y_1-\bar Y_0$
satisfies 
\beq
\delta\bar Y_1 = (1+\bar F\; \bar w)^{-1}\left(\bar C-\bar Y_0-\bar F\; \bar w \;\bar Y_0-\delta(F*Y_0)\right)\;.
\eeq
At the $(n+1)$-th iteration we similarly find 
\beq\label{eq:iterN}
\delta \bar Y_{n+1}=(1+\bar F\; \bar w)^{-1}\left( \bar C-\bar Y_n-\bar F\; \bar w \;\bar Y_n-\delta(F*Y_n)\right)\;.
\eeq
In practice 
the iteration quickly converges to $\delta Y_n=0$, which implies that $Y$ satisfies Eq.~(\ref{eq:dyne}).
\section{DMFT solutions}
DMFT requires 
the solution of the local Green's function $G_{ii}\equiv G  $ from an impurity model with action:
\beq
\mathcal{S}=-i\int_{\mathcal C}dt'\,H_\text{loc}(t') -i \int_{\mathcal C}dt_1dt_2 \Lambda(t_1,t_2) c_\sigma^\dagger(t_1)c_\sigma(t_2)\;,
\eeq 
where $H_\text{loc}$ is the local part of the Hamiltonian, and the hybridization function 
$\Lambda(t_1,t_2)$
is determined self-consistently. 
As stated in the main text, we consider a Bethe lattice in the infinite coordination limit $Z\to\infty$, 
with the $V_{ij}$ corresponding to a semi-elliptic density of states:
\beq
\rho(\epsilon)=\frac{1}{L}\sum_{k}\delta(\epsilon-\epsilon_{k})=\frac{1}{2\pi V}\sqrt{4V^2-\epsilon^{2}}\;,
\eeq 
which allows 
a closed form of the self-consistency: $\Lambda(t_1,t_2)= V^{2}G(t_1,t_2)$ 
\cite{Georges96}.

As in the standard nonequilibrium case, the solution of the nonequilibrium problem, i.e., the evaluation  of the Green's function from the expectation value $G(t_1,t_2) = -i\, {\rm Tr} [\tio e^{-\mathcal{S}}c(t_1)c^\dagger(t_2)]/ {\rm Tr} [\tio e^{-\mathcal{S}}$], is numerically the most challenging part. However, all approaches based on perturbation expansions can be readily rewritten, including numerically exact Quantum Monte Carlo algorithms \cite{Werner2010} (used in the context of the Hubbard model), and closed equations of motion (used in the context of the Falicov-Kimball model)  \cite{Freericks2006}.
\par
Below we illustrate the DMFT solution of the Falicov-Kimball and the Hubbard model.
In both cases the  lattice Hamiltonian has the contour-dependent form:
\beq\label{eq:contdepham}
H(t)=\left\lbrace
\begin{array}{cc}
H&\quad   \text{ if } t\in \mathcal C_1\\
H_0&\quad \text{ if } t\in \mathcal C_2\\
H_0&\quad \text{ if } t\in \mathcal C_3\;,
\end{array}
\right.
\eeq
where 
\beq
H=H_0+U\sum_i n_{i\uparrow}n_{i\downarrow}\;.
\eeq
%
\subsubsection{Exact DMFT solution of the FKM}

The DMFT equations for the Falicov-Kimball model are formally identical to the sudden-quench case~\cite{Eckstein_PRL08,Eckstein2008c}, the only difference being  in 
the explicit contour-dependence of the Hamiltonian. 
However, this difference plays an important role, since in the standard case the problem has also an analytical solution ~\cite{Eckstein_PRL08}, while in our case we have to resort to a numerical solution.
 
The central quantity  for the DMFT solution is the local Green's function $G(t,t')$ of the itinerant fermions, which is a weighted  sum of two components~\cite{Eckstein_PRL08}:
\beq
G(t, t')= w_{0} Q(t, t') + w_{1} R(t, t')\label{eq:Gloc}\;.
\eeq
In particular, the component $Q(t,t')$ describes lattice sites where 
immobile
electrons are absent, while the component 
$R(t,t')$ takes into account the presence of the potential due to the $f$-electrons. 
The weight $w_{1}=1-w_{0}$  is the average number of localized particles and we take $w_{0}=0.5$, i.e. half-filling.
The expectation value of the double occupancy $d_{\rm FK}(t)$ is easily computed within the DMFT formalism from the Green's function:
\beq\label{eq:fc-occ}
d_{\rm FK}(t)=-iw_{1}R^{<,1}(t,t),
\eeq
where the lesser component of $R$ is defined as
\beq
R^{<,1}(t,t)\equiv\lim_{t'\to t^+}R_{11}(t,t'), 
\eeq
keeping in mind the matrix structure (\ref{eq:Gmatrix}).

We briefly recall now the steps of the DMFT self-consistency loop.
Starting from some initial guess for the hybridization $\Delta(t,t')$,
the local Green's function~(\ref{eq:Gloc}) can be determined solving~\cite{Eckstein_PRL08}
\begin{eqnarray}
&\left[i\partial_{t}+\mu\right]Q(t,t')-\Delta*Q (t,t')=\delta_{\mathcal C}(t,t'),\label{eq:fkmQ}\\
&\left[i\partial_{t}+\mu-U(t)\right]R(t,t')-\Delta*R (t,t')=\delta_{\mathcal C}(t,t'),\label{eq:fkmR}
\end{eqnarray}
where $(i\partial_t+\mu)\equiv g_0^{-1}$ is the inverse of the single-particle, noninteracting Green's function
(we similarly define also $(i\partial_t+\mu-U(t))\equiv g_1^{-1}$).
In Eqs.~(\ref{eq:fkmQ}) and (\ref{eq:fkmR})  
$\delta_{\mathcal C}(t,t')$ is the contour delta function satisfying 
\beq
\delta_{\mathcal C}(t,t')=\partial_t\theta_{\mathcal C},\quad\int_{\contour}d\bar t\, \delta_{\contour}(t,\bar t)g(\bar t)=g(t)\;\forall g(t)\;,
\eeq
with the time derivative defined according to the branch:
\beq
\partial_t g(t)=\left\lbrace
\begin{array}{ll}
\partial_t g(t^{\pm}) &t\in \contour_{1,2}\\
i\partial_{\tau}g(-i\tau)& t=-i\tau\in\contour_3\;.
\end{array}
\right.
\eeq
The numerical implementation of the derivative and  the delta function is nontrivial, as discussed in Ref.~\onlinecite{Freericks2008}.
However we avoid this problem because we do not solve directly Eqs.~(\ref{eq:fkmQ}) and (\ref{eq:fkmR}), but rather their integral version:
\begin{eqnarray}
(1-g_0*\Delta)*Q&=g_0,\label{eq:fkm1}\\
(1-g_1*\Delta)*R&=g_1.\label{eq:fkm2}
\end{eqnarray}
The next step is computing the new
hybridazion function from the knowledge of $G(t,t')$. This becomes a simple task if the semi-elliptic density of states 
is assumed~\cite{Eckstein_PRL08} (as we do in the main text): 
\beq\label{eq:fkmDelta} 
\Delta(t,t')=V^{2}G(t,t').
\eeq
wher $V$ is the hopping energy scale.
With this new solution for the hybridization, a new iteration starts by inserting $\Delta$ in Eqs.~(\ref{eq:fkm1}) and ~(\ref{eq:fkm2}). The set of equations ~(\ref{eq:Gloc}), (\ref{eq:fkm1}),  (\ref{eq:fkm2}) and (\ref{eq:fkmDelta}) is solved self-consistently until convergence is reached.
For the results shown in this paper, the convergence criterion is defined 
by
\beq
\varepsilon={\rm max}_{i\in[0,N]} \;\vert \Re(d_{\rm FK}(t_i)^{(n+1)})-\Re(d_{\rm FK}(t_i)^{(n)})\vert,
\eeq
which compares the real part of the double occupancies at the $(n)$-th and the previous DMFT iteration, terminating the computation 
if $\varepsilon <\varepsilon_{\rm max}$. Typical values of $\varepsilon_{\rm max}$ are $\sim 10^{-6}$.

With the self-consistency equations displayed above, the number of iterations $n_{\rm iter}$ necessary for the DMFT to converge is naively expected to increase both with the interaction $U$ and 
with the maximum time $t_{\rm max}$ up to which the evolution is calculated. 
However, in the presence of coexisting solutions, also at short or intermediate times $n_{\rm iter}$ can 
become large. 
A way to accelerate convergence 
is slightly modifying the self-consistency step, so that
the hybridization function at iteration $(n+1)$ is constructed from a combination of the newly-computed Green's function $G^{(n)}(t,t')$ and the $G^{(n-1)}(t,t')$ obtained at the previous step:
\beq\label{eq:combination}
\begin{split}
\Delta^{(n+1)}(t,t')=&V^{2}\left(G^{(n-1)}(t,t')+\right.\\
&\left.\gamma(G^{(n)}(t,t')-G^{(n-1)}(t,t'))\right).
\end{split}
\eeq
The parameter $\gamma$ controls the mixing of the two solutions, $\gamma=1$ corresponding to the case of Eq.~(\ref{eq:fkmDelta}).
\subsubsection{DMFT solution of the Hubbard model with continuous-time Monte Carlo}
A numerically exact solution of the nonequilibrium DMFT equations for the Hubbard model can be obtained with a continuous-time Monte Carlo impurity solver. The weak-coupling approach \cite{Werner2009, Werner2010} 
uses the noninteracting impurity Green's function $G_0$ as an input. This function is related to the hybridization function $\Delta$ by 
\beq\label{eq:G0fromDelta}
G_0^{-1}(t,t')=(i\partial_t+\mu)\delta_\mathcal{C}(t,t')-\Delta(t,t')\;.
\eeq
The quantity measured in the Monte Carlo simulation is the improper self-energy\cite{Werner2010, REVIEW} $X(t_1,t_2)$, which is related to the noninteracting impurity Green's function $G_0$ and self-energy $\Sigma$ by 
\beq\label{eq:opX}
X_\sigma*G_{0,\sigma}=\Sigma_{\sigma}*G_{0,\sigma}.
\eeq
Using the Dyson equation~(\ref{eq:dyson}) and Eq.~(\ref{eq:opX}) we obtain the relation 
\beq\label{eq:self-energy}
(1+X_\sigma*G_{0,\sigma})*\Sigma_{\sigma}=X_{\sigma},
\eeq
which for given $X_\sigma$ can be solved in a stable manner to yield the self-energy.
Once the self-energy is obtained, the DMFT self-consistency step is performed as follows: the new hybridization function  $\Delta$ is found via the Dyson equation~(\ref{eq:dyson}) and the self-consistency equation (\ref{eq:fkmDelta}), and then the new noninteracting Green's function $G_{0}$ results from inverting Eq.~(\ref{eq:G0fromDelta}).
The double occupancy is extracted from the relation
\begin{equation}
U(t)\langle n_\sigma(t)[n_{\bar \sigma}(t)-\tfrac12]\rangle=-i[\Sigma*G_\sigma]^{<}(t,t).
\end{equation}

%

\end{document}